\documentclass[amsmath,amssymb, aps,nature,twocolumn]{revtex4-2}

\usepackage{graphicx}% Include figure files
\usepackage{dcolumn}% Align table columns on decimal point
\usepackage{bm}% bold math
\usepackage{subcaption}
\usepackage[justification=Justified]{caption}

\usepackage{xcolor}
\usepackage{float}
\usepackage{siunitx}
\usepackage{amsmath,amssymb}
\renewcommand{\selectlanguage}[1]{}
\usepackage[final,
	pdftex,
	colorlinks=true,
  	urlcolor=blue,
  	linkcolor=blue,
  	citecolor=blue
]{hyperref}

\usepackage{silence}
\newcommand{\ket}[1]{\ensuremath{\left\vert{#1}\right\rangle}}
\newcommand{\bra}[1]{\ensuremath{\left\langle{#1}\right\vert}}

\begin{document}

\preprint{APS/123-QED}

\title{Many-Body Super- and Subradiance in Ordered Atomic Arrays}

\author{Alec Douglas, Lin Su, Michal Szurek, Robin Groth, Sandra Brandstetter, Ognjen Markovi\'c, Oriol Rubies-Bigorda, Stefan Ostermann, Susanne F. Yelin, Markus Greiner}
 \affiliation{Department of Physics, Harvard University.}
\thanks{amdouglas@g.harvard.edu,\\ mgreiner@g.harvard.edu}

\date{\today}% It is always \today, today,

\begin{abstract}

When quantum emitters couple indistinguishably to light, they can synchronize into a collective light–matter system with radiative properties profoundly different from those of independent particles \cite{dicke_coherence_1954,gross_superradiance_1982}. To date, the resulting collective effects have largely been confined to point-like or homogeneous ensembles \cite{gross_superradiance_1982,chang_colloquium_2018}. Here, we open access to a qualitatively new collective regime by realizing geometrically ordered, spatially extended atom arrays with subwavelength spacing \cite{masson_universality_2022,rubies-bigorda_characterizing_2023,sierra_dicke_2022,bettles_enhanced_2016,shahmoon_cooperative_2017}. This establishes a fundamentally new platform in which collective emission is no longer confined to a single Dicke mode but instead emerges from a ordered network of photon-mediated interactions. We find that 2D atom arrays undergo strong super and subradiant emission. Despite subwavelength spacing, we achieve site-resolved imaging and directly observe the build-up of spatial correlations, demonstrating the transformation of cooperative decay into a strongly correlated many-body process. We observe extensive scaling of superradiance, uncover superradiant revivals, and reveal the ferromagnetic nature of superradiance and the antiferromagnetic nature of subradiance. Our results realize a novel programmable platform for exploring and utilizing dissipative many-body quantum physics \cite{noh_quantum_2016, chang_quantum_2014}, opening new possibilities for photon capture, storage \cite{facchinetti_storing_2016, asenjo-garcia_exponential_2017,rubies-bigorda_photon_2022}, and atom–photon entanglement \cite{bekenstein_quantum_2020}.

\end{abstract}

\maketitle

Collective light-matter interactions are among the most striking manifestation of many-body quantum optics. When many atoms radiate together, they can develop collective coherence, producing intense bursts of superradiant light, or, conversely, arrange into dark subradiant modes that strongly suppress decay.~\cite{dicke_coherence_1954,rehler_superradiance_1971,lehmberg_radiation_1970,bonifacio_quantum_1971,yukalov_cooperative_2000}. Traditionally, experiments have accessed cooperative regimes by approaching idealized point-like systems, either by densely packing emitters within the spatial extent of a radiative wavelength~\cite{raimond_collective_1982,inouye_superradiant_1999,facchinetti_storing_2016,guerin_subradiance_2016} or  engineering cavity atom-light couplings~\cite{slama_superradiant_2007,baumann_dicke_2010,bohnet_steady-state_2012}. 
Consequently, experimental studies of cooperative phenomena have largely treated the system as a uniform, homogeneous ensemble, confining the physics to the semiclassical regime and precluding applications in quantum-photonics.

The situation changes fundamentally when emitters are instead arranged in a controllable sub-wavelength array \cite{masson_universality_2022}. In this regime, the radiative couplings between atoms form a long-range, anisotropic interaction network, creating a dissipative, strongly correlated quantum system that can be programmed, and where the regular subwavelength spacing prevents lossy scattering events. Unlocking this regime has been a long-standing goal in quantum optics with recent experimental effort in this direction, such as realizing atomically thin mirrors \cite{rui_subradiant_2020}. 
It has been theoretically proposed that sub-wavelength arrays generate a rich spectrum of super- and subradiant modes far beyond the semi-classical Dicke limit. Crucially, however, such a system would provide a new  platform for programming complex light-matter systems entirely without cavities or nanophotonic structures. Subradiant modes can act as long-lived photon storage channels \cite{asenjo-garcia_exponential_2017,rubies-bigorda_photon_2022,manzoni_optimization_2018,ballantine_quantum_2021}, while superradiant modes enable fast and directional photon release \cite{masson_many-body_2020,shahmoon_cooperative_2017,bettles_enhanced_2016,rubies-bigorda_photon_2022,ballantine_quantum_2021}. So far, however, populating such modes in subwavelength arrays has remained experimentally elusive.

\begin{figure*}
    \includegraphics[width=\textwidth]{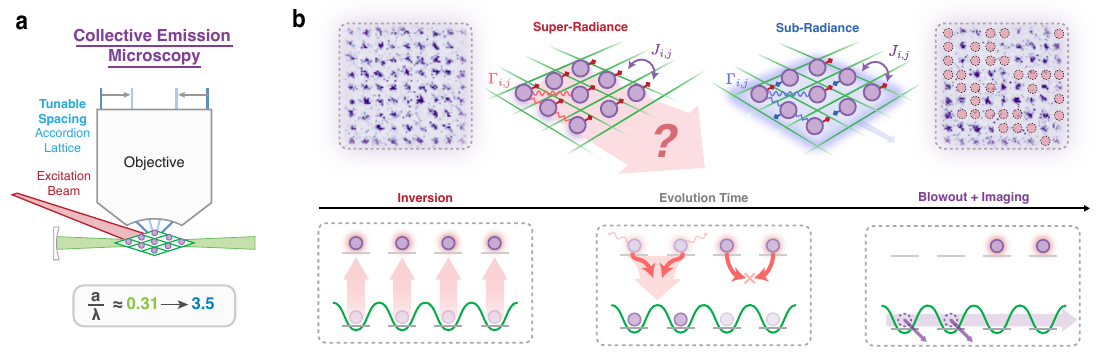}
    \caption{
    \textbf{Subwavelength atom arrays - a new, site resolved platform for many-body quantum optics }\ 
    \textbf{(a.)} 
    An array of erbium atoms with a small spacing of $\qty{266}{\nano\meter}$ is created by loading ultracold 168Er atoms into a counter-propagating optical lattice. The atomic spacing is significantly smaller than the $\qty{841}{\nano\meter}$ atomic transition used in this work, resulting in a minimum spacing to wavelength ratio of  $a/\lambda =0.316$. At this short distance, the dipole moments of the excited atoms interfere in the near field, allowing for cooperative emission events, $\Gamma_{i,j}$, where the spontaneously emitted light can interfere constructively (destructively), leading to collective super (sub) radiance. Additionally, evanescent photons give rise to coherent spin exchange interactions between nearby atoms, $J_{i,j}$.
    We can transfer the atoms into a tunable spacing optical lattice, produced by projecting $\qty{488}{\nano\meter}$ light (blue) through a high numerical aperture objective. The spacing of this accordion lattice is tunable from $\qty{266}{\nano\meter}$ to $\qty{3}{\micro\meter}$, widely varying the ratio of atom spacing and emission wavelength. \textbf{(b.)} The regular atom array is prepared by creating a $>$98\% filling Mott insulator in either our low disorder $\qty{532}{\nano\meter}$ lattice, or our tunable spacing $\qty{488}{\nano\meter}$ accordion lattice. We then apply a global inversion pulse and allow the system to evolve for $t_e$ so that cooperative emission may occur. We image the excited state population by rapidly blowing out all ground state atoms in $0.02\tau$ and expanding the lattice spacing to $\qty{3}{\micro\meter}$ for imaging with single-site resolution. 
    }
    \label{fig:setup}
\end{figure*}

Here, we experimentally realize such a sub-wavelength array and study this novel quantum regime for the first time. We create virtually perfect sub-wavelength arrays by confining ultracold atoms in an optical lattice with lattice spacing well below the emission wavelength. Quantum gas microscopy further allows atoms to be detected individually while the array emits collectively. This capability---simultaneously achieving subwavelength spacing and site-resolved imaging---opens access to this strongly correlated regime that has long been theorized but was not previously accessible. We study how cooperative emission builds and decays locally, how spatial correlations evolve during the process, and how extended arrays explore a rich landscape of super- and subradiant modes. Specifically, we discover the emergence of ferromagnetic spin textures upon initial emission, followed by antiferromagnetic spin textures and dramatically slowed decay arising from interference between multiple subradiant branches. This directly demonstrated the strongly correlated quantum nature of the newly accessible regime and sets the stage for quantum photonics in this novel platform.

\section{Sub-wavelength atom arrays}

To assemble a regular sub-wavelength array of identical quantum emitters, we trap ultracold Erbium atoms in an optical lattice.  By loading a single sheet of a three-dimensional lattice, we create a two-dimensional (2D) array of atoms. The lattice spacing can be as small as 266nm, much less than half the wavelength $\lambda=\qty{841}{nm}$ of the optical transition, eliminating any Bragg modes. An accordion lattice with tunable spacing between 266\,nm and $\qty{3}{\micro\meter}$ allows us to continously tune the array distance. To localize each atom to well below the optical wavelength and suppress photon recoil, the atoms are tightly confined on each lattice site and cooled to their motional ground state, such that their zero-point motion is suppressed to $\sqrt[3]{l_x l_y l_z} / \lambda \approx 0.038$ (where $l_x$ is the harmonic oscillator length in the x direction). We create near-unity ($>$98\%) filling of up to 1000 atoms in a 2D arrangement by creating a Mott insulator from a Bose-Einstein condensate. The tunable spacing accordion lattice enables us to "zoom in" and resolve and image each subwavelength lattice site with high fidelity by increasing the lattice spacing just before imaging ~\cite{su_fast_2024}.

The atoms are excited on an effective two-level system by driving the narrow \qty{8}{kHz} transition at \qty{841}{\nano\meter} with \(\sigma^-\)-polarized light, coupling the stretched ground state \(|g\rangle = |J = 6, m_J = -6\rangle\) to the excited state \(|e\rangle = |J = 7, m_J = -7\rangle\). We track the evolution of the many-body quantum system with single-site imaging, which allows direct measurement of which atoms are in the excited state. Following a collective excitation and evolution, we remove all ground-state atoms using a resonant 401\,nm beam, leaving only excited atoms to decay and be imaged (Fig. \ref{fig:setup} b.). This method circumvents the inefficiencies of collecting emitted photons, particularly problematic for subradiant modes with non-directional emission profiles, and avoids the noise limitations of single-photon detectors, such as dark counts. Instead, by imaging the residual excited atoms, we access the population and spatial coherence of collective modes with high fidelity \cite{glicenstein_probing_2025}. This direct readout enables detailed characterization of the system and provides new insights into the many-body dynamics of radiative systems, including spatial and phase correlations that remain hidden in traditional photon-counting approaches.

\begin{figure*}
    \includegraphics[width=\textwidth]{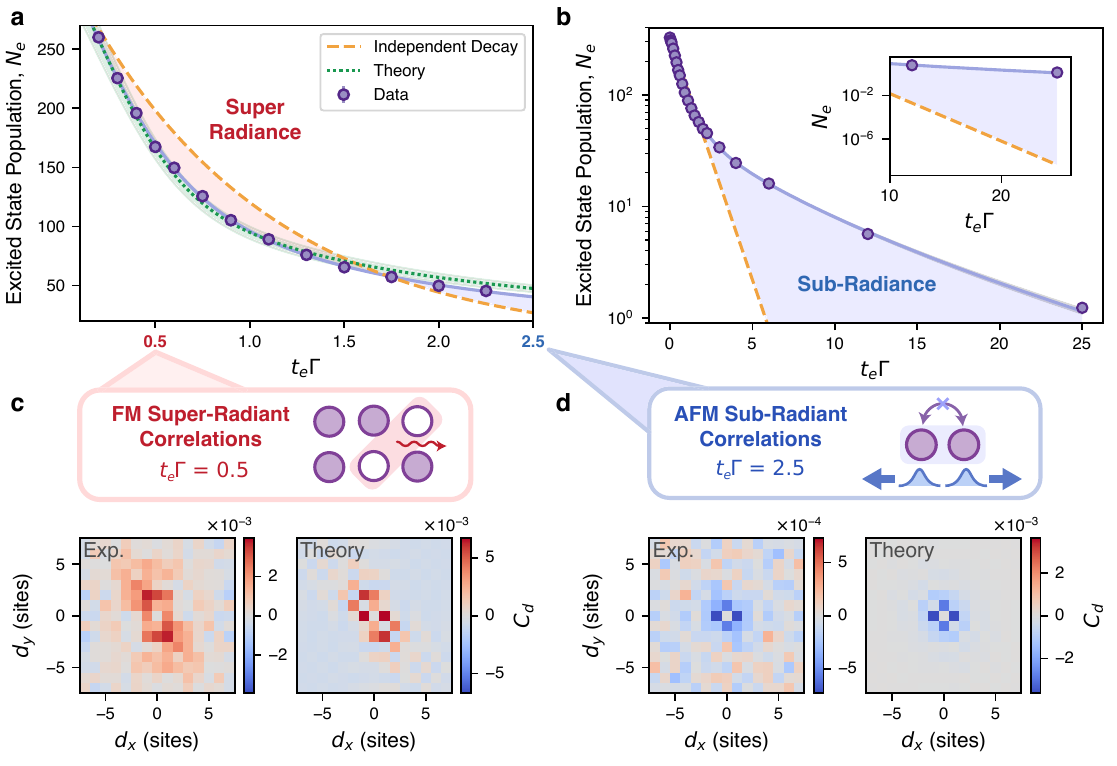}
    \caption{
    \textbf{Direct observation of many-body super- and subradiance in an atomic array}
    \ \textbf{(a.)} Initial decay of the excited state population (purple circles) showing marked deviation from normal exponential decay (orange dashed line throughout the paper). Superradiance is observed as a lower excited state number than the independent decay curve. Theory calculations of the cooperative response (green dotted line with $\sigma$ uncertainty shading) closely match the measured population trace. 
    \textbf{(b.)} At long times, we observe a substantially larger excited state population than independent decay would suggest, indicating a dynamical population of dark states, or subradiance. The inset zooms in on the last two data points, highlighting deviation from independent decay by many orders of magnitude. 
    \textbf{(c.)} (\textbf{(d.)}) The quantum gas microscope allows us to measure the connected density-density correlation $C_\mathbf{d}$ at time 0.5 (2.5) $\tau$. The system of 500 atoms begins fully inverted. We use a linear colormap, with red (blue) indicating ferromagnetic (FM) and antiferromagnetic (AFM) fluctuations. The ferromagnetic correlations reflect correlated decay due to the complicated near-field dipole-dipole Hamiltonian. The antiferromagnetic correlations reflect scattering events and thus correlations between neighbouring subradiant excitations. We see good qualitative agreement with theory (right) computed from experimentally measured fillings and independently measured quantization axis.
    }
    \label{fig:supsub}
\end{figure*}
\section{Super- and Subradiance}
Whether superradiance should exist in an extended, ordered array is not obvious \emph{a priori}. In two dimensions, most optical modes---including the one a fully inverted system initially occupies---radiate predominantly out of the plane, stimulating cooperative effects only among atoms within roughly a wavelength. Under this reasoning, one might expect each subwavelength ``patch'' of the array to behave largely independently, producing a fixed amount of superradiance with no scaling or long-range buildup. Yet a competing intuition points the other way: photons that propagate \emph{along} the array interact most strongly with the emitters, and such guided-like propagation can reinforce itself, amplifying correlations that extend over many lattice sites. This mechanism---a cascade in which long-wavelength spin waves feed emission that in turn reinforces those same modes---has no analogue in the Dicke limit, where symmetry restricts emission to a single collective mode. Whether such a cascade occurs in an extended array, generating gain in long-momentum modes that co-propagate across the system, has remained an open question with recent theoretical work predicting superradiant bursts \cite{masson_universality_2022}. With single-site resolution, we can now directly probe these correlations and determine how superradiance actually develops in two dimensions.

We probe the existence of super and subradiance by preparing the atoms in the $\qty{266}{\nano\meter}$ (\qty{.32}{\lambda}) spacing array in the excited state and  measuring the excited-state population $N_e(t)$ as a function of evolution time (Fig.~\ref{fig:supsub}a). The many-body dynamics are initiated by transferring the atoms to the excited state within a time $\sim \tau/18$, where $\tau = 1/\Gamma = \qty{20}{\micro\second}$ is the independently measured spontaneous lifetime of the excited state (see Methods). We observe that the resulting decay departs rapidly from the exponential profile $N_e(t)\propto e^{-t\Gamma}$ expected for independent atoms (orange dashed line). At early times, the decay is faster, revealing a superradiant enhancement; at late times it becomes slower, revealing the emergence of subradiant states. We developed approximate microscopic numerical simulations that incorporate both coherent and dissipative couplings and known experimental imperfections, and find good agreement (green dashed line; see Methods and Ref.~\cite{rubies-bigorda_characterizing_2023}). To verify that this behaviour stems from cooperative interactions, we repeat the measurement in an array with large spacing ($\qty{3}{\micro\meter}$) far exceeding the emission wavelength. The population now follows a simple exponential decay, confirming the absence of collective effects in the  large spacing limit (see Methods).

To determine whether the superradiant enhancement is driven by correlation buildup, we measure the evolution of spatial correlations between optical excitations with single-site resolution using our quantum gas microscope. Specifically, we evaluate the connected density--density correlation function,
\begin{equation}
C_{\mathbf{d}}
= \frac{4}{N_{\mathbf{d}}}
\sum_{\mathbf{d}=\mathbf{d}_{i,j}}
\left(
\langle \hat{n}_i \hat{n}_j \rangle 
- 
\langle \hat{n}_i \rangle \langle \hat{n}_j \rangle
\right),
\end{equation}
during different regimes of the decay process (Fig.~\ref{fig:supsub}c). At early times, we observe pronounced ferromagnetic (red) correlations of the ``holes'' (decayed atoms) arising from collective emission. Their short-range ($\sim\lambda$ or 3 site distance) structure mirrors the dissipative coupling matrix $\Gamma_{i,j}$, while long-range correlations extend along one lattice axis, consistent with a cascade of long-momentum spin waves propagating across the array. The anisotropy in choice of axis arises from our choice of circular polarization and a $30^{\circ}$ in-plane quantization axis. These correlations peak precisely at the maximum emission rate, providing direct evidence that superradiance in extended arrays is built through the rapid assembly of long-range coherence, rather than purely through local stimulation.

At late times ($t_e \gg \tau$), the dynamics undergo a qualitative transformation: the array settles into a set of highly subradiant, long-lived many-body states in which the photon emission rate is suppressed by an order of magnitude (Fig.~\ref{fig:supsub}b). This behaviour emerges naturally from the system’s own dynamics. During the intermediate regime, the absence of strong symmetries allows population to explore a broad manifold of collective modes. As emission proceeds, the brightest modes decay away first, and the array progressively filters itself into the most subradiant configurations available. This dynamical selection results in a monotonic slowdown of the decay rate and reveals that the long-time behaviour of the system is governed not by the initial inversion, but by the collective dark states it self-organizes into \cite{rubies-bigorda_dynamic_2023}.

With bright modes depleted, the spatial correlations are dictated by the surviving excitations. These correlations develop into clear antiferromagnetic patterns, demonstrating spatial antibunching among the remaining excitations. This behaviour is particularly striking: because each atom can host at most one excitation, any spatial overlap between multi-excitation subradiant modes generates bright components that decay rapidly. The surviving dark excitations therefore exhibit an effective radiative repulsion, reorganizing themselves into patterns that minimize their coupling to free space. Similar effects were predicted in one-dimensional geometries~\cite{asenjo-garcia_exponential_2017,henriet_critical_2019}. The emergence of antiferromagnetic textures highlights that the long-lived subradiant states are strongly correlated many-body states, not simply long-lived single-particle modes.

%Interestingly, around $t \approx 1.6\tau$ the system shows negligible density--density (ZZ) correlations (see Methods), yet the antiferromagnetic XY correlations characteristic of subradiant coherence are already well established. This indicates that the dark-state structure first appears in the phase sector before becoming visible in the density sector, revealing a two-step process by which the array assembles its long-lived, radiatively protected many-body states.

\begin{figure}[htp!]
    \centering
    \includegraphics[width=0.5\textwidth]{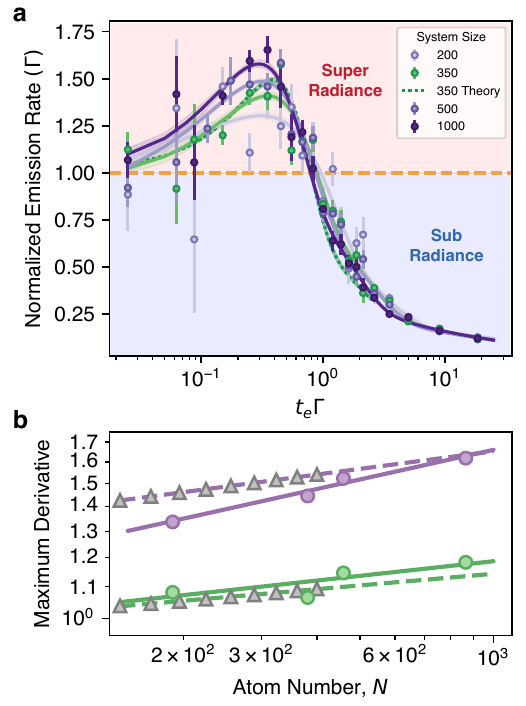}
    \caption{
    \textbf{Showing extensive scaling of superadiance in a mesoscopic atomic array}
    \ \textbf{(a.)} We compare the emission rate with the normalized (flat orange line) exponential decay rate of individual emitters. We observe that the emission rate scales with atom number, showing cooperative emission across the extended array. We extract the emission rate from measured population curves at different atom numbers (circles), normalized to independent decay for the given number of atoms. Solid lines are extracted from numerical derivatives of the population fits with bootstrapped uncertainties.
    \textbf{(b.)} We extract the polynomial scaling exponent of the emission rate with atom number. Purple data points show the maximum emission rate normalized to the number of exited state atoms remaining in the system, while green data points show the maximum emission rate normalized by the number of atoms in the whole array; values greater than one therefore indicate a ``superradiant burst". Solid lines denote linear fits, indicating a power-law relationship between system size and maximum emission rate. Grey triangles (dashed lines) show simulation data points (fits).
    }
    \label{fig:scaling}
\end{figure}

In a cooperative system, the relevant timescale becomes the time-dependent radiative strength of the collective state. For independent emitters, the photon emission rate at time t, $N_e'(t) = -\frac{d N_e(t)}{dt}$, decays exponentially. Dividing this emission rate by the number of excitations at time $t$ returns a constant, $\gamma_0$, which we refer to as the normalized emission or decay rate. For cooperative systems, this normalized metric is no longer constant but instead becomes time-dependent: \begin{math}
    \gamma(t) = -\frac{1}{N_e(t)}\frac{d N_e(t)}{dt}
\end{math}. We use this time-dependent normalized emission (or decay) rate as a metric for collective effects, where $\gamma(t)>\gamma_0$ ($\gamma(t)<\gamma_0$) indicates superradiance (subradiance).

Fig. \ref{fig:scaling}a., reveals that our array begins exactly where an uncorrelated system must: immediately after inversion $\gamma(0)=\gamma_0$. At this moment the state is a product state with no coherences between atoms, and its emission is driven purely by vacuum fluctuations. Decay events seed correlations across the array, and $\gamma(t)$ rises sharply. This early growth of the decay rate signals the emergence of a bright collective mode—a many-body organization that enhances emission. The subsequent sharp peak in $\gamma(t)$ is the quantitative signature of superradiance in an extended, multimode environment. If superradiance were simply the sum of many independent subwavelength patches, the normalized rate 
$\gamma(t)$ would saturate as the array grows. Instead, we observe a clear power-law scaling of the superradiant peak with atom number, $\gamma_{\text{max}}\propto N^\alpha$ with $\alpha_{\text{norm}}=$\qty{1.13(.01)} (Fig \ref{fig:scaling}b.) demonstrating that cooperativity strengthens as atoms are added—even when they lie many wavelengths apart \cite{mok_universal_2024}. This confirms that the array emits as a single, extended object, not as a mosaic of independent regions.

A closely related observable, the peak emission rate per atom $N_e'(t)/N$, exhibits similar behaviour with exponent $\alpha_{\text{raw}}=$\qty{1.063(.015)}. This raw scaling is in excellent agreement with numerical predictions for unity-filled square arrays, while the slightly larger exponent for the normalized rate suggests a crossover in scaling for even larger systems.

In contrast to this size-dependent enhancement at early times, the degree of subradiance at intermediate times depends only weakly on atom number and is consistently overestimated by approximate simulations, likely due to the difficulty of capturing higher-order correlators essential to multi-excitation dark states. At very late times, however, all arrays converge to a common decay floor, $\gamma(t \to \infty) \approx \gamma_0/10$ as shown in Fig.~\ref{fig:scaling}a. This saturation is set not by cooperative physics but by the finite spatial extent of the atomic wave packets---approximately $0.05\lambda$ in the lattice plane and $0.1\lambda$ along $z$ (see Methods)---which ultimately limits how dark the most subradiant modes can become \cite{rubies-bigorda_characterizing_2023}.

Taken together, these results reveal a clear physical mechanism: as the array grows, the superradiant peak becomes a genuinely many-body effect that strengthens with system size, while the long-time subradiant floor becomes a size-independent feature set by microscopic localization. Between these extremes, the normalized emission rate $\gamma(t)$ provides a window into how an extended quantum system builds, amplifies, and ultimately hides its excitations among a hierarchy of collective radiative modes.

\section{Beyond the Dicke Limit}

\begin{figure*}
    \includegraphics[width=\textwidth]{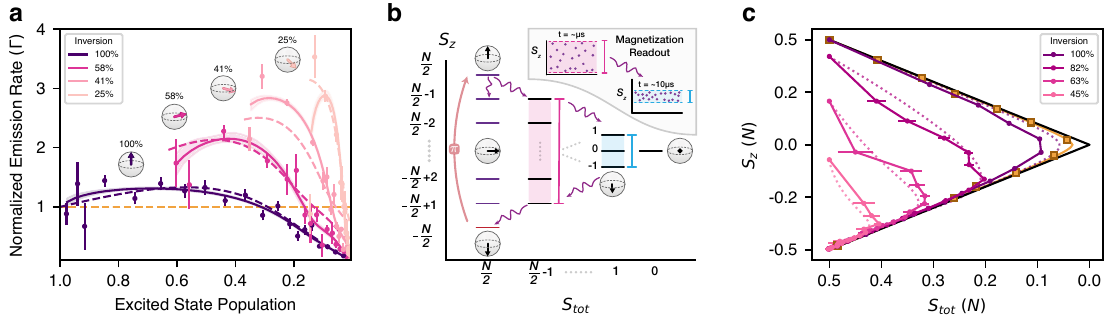}
    \caption{\textbf{Seeding the superradiant burst and exploring beyond the Dicke manifold}
        \ \textbf{(a.)} The emission rate per excited atom drastically increases as we seed the superradiant burst with varying initial excitation fractions. Colored (dashed) lines represent fits to experimental data (theory) with 1 standard deviation uncertainty shading from bootstrap. The decay rate increases as the initial excitation fraction decreases, reflecting the role of laser-induced initial coherences, while a significant portion of the population remains trapped in subradiant states across all excitation fractions. The coherences provide strongly directional single photon emission along the incident laser direction opening the direction of photon capture and release \cite{masson_many-body_2020}.
        \textbf{(b.) (c.)} We characterize and plot the good quantum numbers of Dicke superradiance; angular momentum eigenstates of $N$ spin-$\tfrac{1}{2}$ in terms of their total angular momentum $S_{\text{tot}}$ and their $S_z$ projection. In the Dicke limit---where the atoms interact with a single coherent optical mode---$S_{\text{tot}}$ is conserved, limiting initial states to the $S_{\text{tot}} = N/2$ manifold. However, the emergence of multiple collective decay channels in extended atomic arrays allows the system to traverse the manifold each time an excitation is emitted, as indicated by purple arrows. We measure the average $S_{\text{tot}}$ of the state at time $t$ by applying a $S_{\text{tot}}$-conserving global rotation and measuring the variance, or size, of the coupled spin ladder.
        }
    \label{fig:inversion}
\end{figure*}

When the system is not in the fully excited phase, textures can be imprinted into the atomic array, creating directional emission-- a regime essential for future single-photon catch-and-release, where the system acts as a receiver or transmitter. In the Dicke model, reducing the inversion simply moves the system down a fixed collective ladder without altering the underlying mode structure. In an extended array, however, the initial excitation fraction determines \emph{which} collective modes are seeded, and therefore which radiative pathways the system can explore.

To probe this, we prepare partially inverted states by shortening the excitation pulse. In this regime, the laser no longer excites all atoms uniformly; instead, it imprints a coherent spin wave with a phase gradient along its propagation direction. This pre-seeds the array into a strongly emissive and directional collective mode, and as a result, the initial decay rate per excitation increases nearly linearly as the inversion fraction decreases (Fig.~\ref{fig:inversion}a). Fewer excitations radiate faster because they occupy a more coherent, more directional superradiant mode.

In the limit of a single photon excitation, we expect perfect superradiance. However, for the finite densities we probe, we observe strong non-linear effects. Regardless of the initial excitation fraction, a substantial portion of the population -- exceeding $15\%$ in all cases -- survives into the late-time subradiant tail. This survival cannot arise from the initial overlap with dark modes, which vanishes as the inversion fraction decreases. Instead, the subradiant population is generated dynamically: as cooperative emission rapidly drains the brightest components, the remaining excitations are funneled into long-lived dark configurations due to interexcitation interactions, potentially allowing for the creation of well-defined photon Fock states.

Partial inversion also allows us to track how the array navigates its exponentially large many-body Hilbert space during decay. We measure the transverse magnetization $M^2 = \langle S_x^2 + S_y^2 \rangle$ by applying global $\pi/2$ pulses with random phases and combine it with $S_z$ to reconstruct the effective total spin, $S_{\mathrm{tot}} = \sqrt{M^2 + S_z^2}$, for different inversion fractions (Fig.~\ref{fig:inversion}c). The resulting trajectories reveal a consistent pattern: the system begins in high-spin, bright configurations and then migrates toward low-spin sectors associated with strongly subradiant modes. The symmetry expected at $50\%$ inversion in both the Dicke limit and spontaneous decay is absent here (see Methods), replaced by an asymmetry that signals the presence of multiple radiative channels unique to extended arrays. For small excitation angles, the laser seeds a sharply defined superradiant spin wave, forcing nearly all emission into this bright channel; as the system decays and $S_z$ decreases, it inevitably moves into the lowest-spin manifolds compatible with the remaining excitations. The late-time states consistently lie near the minimal total spin allowed, confirming that the surviving excitations occupy subradiant modes with wavevectors markedly different from those of the initial drive and with largely suppressed coupling to free space.

\begin{figure}[htp!]
    \includegraphics[width=0.5\textwidth]{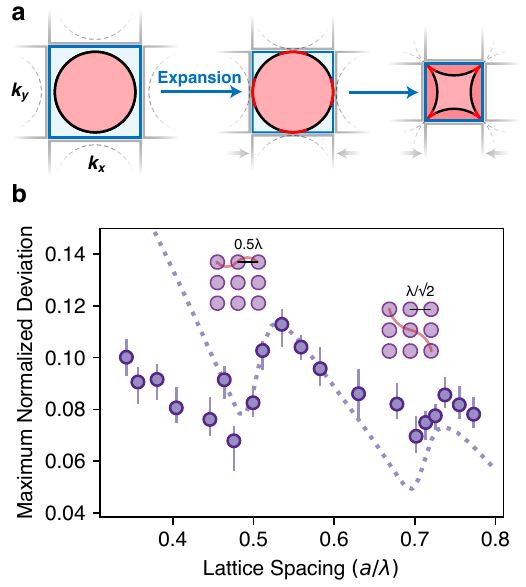}
    \caption{\textbf{Superradiant geometric resonances in an ordered array}. 
    \ \textbf{(a.)} 
    Light has an intrinsic maximum in-plane momentum which, for a subwavelength array, is smaller than the Brillouin zone of the periodic atomic lattice. This mismatch creates two distinct regions: a superradiant region (red), where atoms can emit photons into free space, and a subradiant region (blue), where emission is forbidden.
    At the depicted lattice spacings, the allowed photon momentum extends to the next Brillouin zone, folding over and adding new allowed scattering channels. \textbf{(b.)} We expand our accordion lattice prior to inverting the system to probe how different lattice spacings affect superradiance. Measuring population traces at each different spacing, such as in Fig. \ref{fig:supsub}a., we plot the maximum deviation from independent decay. Purple dots represent experimental data, and dotted lines represent theory with an offset extracted from a least squares fit. We observe geometric resonances corresponding to interatomic distances commensurate with the excitation wavelength at $a/\lambda=0.5$ and $1/\sqrt{2}$, as indicated by the inset sketches.
    }
    \label{fig:spacing}
\end{figure}
\section{Engineering the Vacuum Coupling}
\label{section: variable_spacing}

Varying the geometry profoundly modifies how an extended array couples to free space by changing the set of allowed photonic momenta-the regular wavelength array eliminates incoherent scattering, allowing a lossless atom-photon interface. When the system is an elongated disordered cloud with directional emission, incoherent Rayleigh scattering dominates and photons leak into the vast continuum of free-space modes. By contrast, a regular array enforces strict momentum conservation: scattering becomes coherent Bragg diffraction, and photons exchange momentum with the lattice rather than diffusing incoherently. By taking the inter-emitter spacing to $a<\lambda/2$ the momentum transferred becomes larger than the allowed photon momentum completely suppressing scattering processes. Unlike optical cavities or waveguides, where optical mode confinement is imposed by mirrors or structures, here the geometry itself acts as a spectral filter—selectively permitting or forbidding entire families of collective radiative modes and eliminating lossy scattering pathways.

To investigate this, we use our tunable-period accordion lattice to vary the interatomic spacing while preserving the underlying atomic pattern. Beginning in the deeply subwavelength regime $a/\lambda = 0.316$, we expand the lattice isotropically up to $a/\lambda \approx 0.8$. For each spacing, we prepare a fully inverted array and measure $N_e(t)$, extracting the maximum deviation from independent exponential decay (Figs.~2a and 5b). This metric tracks the strength of the superradiant burst and directly reflects how efficiently the array channels population into bright collective modes.

We observe that rather than decreasing smoothly with spacing, the cooperative enhancement exhibits a pronounced oscillatory structure, with sharp peaks when the spacing becomes commensurate with specific fractions of the optical wavelength—most prominently at $a = \lambda/2$ and $a = \lambda/\sqrt{2}$. These resonances provide clear evidence that new radiative pathways open discretely as the spacing increases. At these special geometries, the maximum allowed in-plane photonic momentum reaches the edge of the Brillouin zone defined by the array. Spin waves that previously lay outside the light cone are folded back into it through Bragg scattering enabling efficient radiation. Because these diffraction orders are indistinguishable within the array, they interfere coherently and enhance emission.

\section*{Conclusion}
We have realized collective light-matter interactions in cavity-free two-dimensional atomic arrays, entering a regime where radiative dynamics are no longer dictated by the Dicke limit of a single mode but emerge from structured, photon-mediated many-body interactions. By leveraging the single-site resolution of a quantum gas microscope, we directly visualized a transition from ferromagnetic superradiant bursts to antiferromagnetic subradiant states. These many-body states act as emergent, photon-trapping configurations, enabling a powerful form of dissipative coherence. Precise control of the many-body states demonstrated here establishes a pathway toward tunable quantum-coherent atom-photon interfaces to propel both fundamental science and quantum photonic devices. 

Our platform unlocks new avenues for fundamental science. The interplay between coherent many-body dynamics and structured dissipation offers a rich testing ground for open quantum systems, non-Hermitian physics, and quantum measurement theory. For instance, optical pumping on cooperative transitions can be interpreted as a form of weak measurement, providing a route to measurement-induced entanglement and state purification in strongly correlated systems. This raises the prospect of using engineered collective decay not merely as a feature to be suppressed, but as a resource for preparing and stabilizing exotic quantum phases and entanglement~\cite{kastoryano_dissipative_2011,reiter_scalable_2016,wetter_observation_2023,kamar_markovian_2025, lin_dissipative_2025,zhan_rapid_2025}, such as protected topological edge modes~\cite{perczel_topological_2017}. Further understanding of cooperative effects is essential for the next generation of optical lattice clocks, where collective dipole shifts perturb the interrogated resonance~\cite{hutson_observation_2024} and subradiant states offer extended coherence times, expanding the atomic species available for optical clocks~\cite{henriet_critical_2019}.

The programmable control over emission pathways and radiative lifetimes demonstrated here lays the foundations for a new class of quantum photonic devices. Applications include photon sources with programmable directionality via single-site phase imprinting~\cite{clemens_collective_2003,masson_many-body_2020}, and lossless optical memory that allows for photon storage and retrieval by interchanging between super- and subradiant modes~\cite{rubies-bigorda_photon_2022,asenjo-garcia_exponential_2017,manzoni_optimization_2018,ferioli_storage_2021,cech_dispersionless_2023,facchinetti_storing_2016,ballantine_quantum_2021}. Furthermore, the platform enables on-demand generation of entangled photons—either through extrinsic entangling interactions or intrinsic interactions between trapped photonic excitations~\cite{santos_generating_2022,srakaew_subwavelength_2023,bekenstein_quantum_2020,rubies-bigorda_deterministic_2025}—providing the key ingredients for quantum communication and distributed quantum computing.

\begin{acknowledgments}
We thank Ana Asenjo-Garcia and Darrick Chang for their insightful conversations on subradiant fermionization and Umklapp scattering. We thank Ignacio Cirac and Francesca Ferlaino for insightful discussions, and Lev Kendrick, Aaron Young, Simon Hollerith, Yidan Wang, and Brandon Grinkemeyer for careful readings of this manuscript.

\textbf{Funding:}  We are supported by U.S. Department of Energy Quantum Systems Accelerator DE-AC02-05CH11231, National Science Foundation Center for Ultracold Atoms PHY-2317134, National Science Foundation PHY-2207972, Army Research Office Defense University Research Instrumentation Program W911NF20-1-0104, Office of Naval Research Vannevar Bush Faculty Fellowship N00014-18-1-2863, Defense Advanced Research Projects Agency Optimization with Noisy Intermediate-Scale Quantum Devices W911NF-20-1-0021, Army Research Office Grant W911NF20-1-0163, and Gordon and Betty Moore Foundation Grant GBMF11521. A.D. acknowledges support from the NSF Graduate Research Fellowship Program Grant DGE2140743. O.R.-B. acknowledges support from Fundación Mauricio y Carlota Botton and from Fundació Bancaria “la Caixa” (LCF/BQ/AA18/11680093). The computations in this paper were run in part on the FASRC Cannon cluster supported by the FAS Division of Science Research Computing Group at Harvard University.

\textbf{Author contributions:} A.D., L.S., and M.S. ran experiments and analyzed experimental data. A.D., L.S., M.S., R.G., and O.M. contributed to the experimental apparatus. A.D. conceived the experiment. O.R.-B., S.O., O.M., and R.G. performed numerical simulations. M.G. and S.Y. supervised all works. All authors discussed the results and contributed to the manuscript.

\textbf{Competing interests:} M.G. is a cofounder, shareholder, and consultant of QuEra Computing. S.F.Y. is also a shareholder of QuEra Computing. All other authors declare no competing interests.
\end{acknowledgments}

\section{Methods}

\subsection{Mott Insulator Preparation and Differential Light Shift}
We first create a Bose-Einstein Condensate in less than a second as described in~\cite{phelps_sub-second_2020}, load it into a single vertical layer, and then adiabatically load a near unity ($>98\%$ filling) Mott insulator as described in~\cite{su_dipolar_2023}. The Mott insulator is initially loaded into a \qty{1064}{nm}-wavelength retro-reflected lattice in the vertical direction and \qty{532}{nm} wavelength lattices in the x-y plane. We use a DMD during loading to shape the Mott insulator into a square with sharp ($<$2 sites) edges. We then optionally transfer to a \qty{488}{nm} wavelength tunable accordion lattice to vary the system's inter-atomic distance. Each of these lattices has different trap depths and differential light shifts, and the accordion lattice has different depths and trap frequencies vs spacing due to the lattice beams having some finite clipping on the objective. We compile these parameters in Table \ref{tab:traps}. The table includes lattice depths, the measured differential Stark shifts, and the Lamb-Dicke parameter $\bar{\eta}^2=(\eta_g^2+\eta_e^2)/2$ averaged between the two states. Ground-state trap frequencies were measured using lattice modulation spectroscopy. The differential light shift was measured by sweeping lattice depth and measuring the change in transition frequency for each lattice.

\begin{table}[b]%The best place to locate the table environment is directly after its first reference in text
\caption{\label{tab:table1}%
Lattice Parameters
}
\begin{ruledtabular}
\begin{tabular}{ccccccc}
\textrm{Lattice}&
$U_g (kHz)$&
$U_e/U_g$&
$\nu_g (kHz)$&
$1-p_{\text{excite}}$ &
$\bar{\eta}^2$\\
\colrule
$\lambda_{1064}^z$  & 97 & 0.8 & 18 & 0.002 & 0.098\\
$\lambda_{532}^x$  & 93 & 0.69 & 39.5 & 0.004 & 0.046\\
$\lambda_{532}^y$  & 113 & 0.69 & 43.6 & 0.004 & 0.042\\
$\lambda_{488}^{x}, a=266$ & 95.7 & 0.56 & 40 & 0.010 & 0.048\\
$a=300$ & 173 & & 54 & & 0.036\\
$a=350$ & 183 & & 56 & & 0.034\\
$a=400$ & 184 & & 56 & & 0.034\\
$a=841$ & 194 & & 57 & & 0.034\\
$a=2000$ & 150 & & 50 & & 0.038\\
$\lambda_{488}^{y}, a=266$ & 156 & 0.56 & 51 & 0.010 & 0.038\\
$a=300$ & 281 & & 69 & & 0.028\\
$a=350$ & 297 & & 71 & & 0.027\\
$a=400$ & 298 & & 71 & & 0.027\\
$a=841$ & 314 & & 73 & & 0.026\\
$a=2000$ & 244 & & 64 & & 0.030\\
\end{tabular}
\label{tab:traps}
\end{ruledtabular}
\end{table}

\subsection{Rabi Drive}
We drive the atoms resonantly to the $\qty{841}{nm}$ excited state. To isolate an effective two-level system along the $\sigma^-$ transition, we apply a \qty{.34}{Gauss} magnetic field roughly $+30$ degrees from the x-axis, splitting the $\pi$ transition by \qty{550}{kHz}. Furthermore, we apply the excitation beam roughly along the magnetic field, $+15$ degrees from the x-axis, with circular polarisation. We estimate that after full inversion, the excited state population mistakenly transferred into the $\pi$ transition is less than $0.003$. 

Throughout this work, the $\sigma^-$ transition Rabi frequency is kept at roughly $55\Gamma$ = \qty{440}{kHz}. The optimal Rabi time drifts between $1.1$ and $1.2$ us, which implies an excited state population of \qty{0.96\pm 0.01}, when accounting for the approximately 4 times enhanced normalized emission rate as measured in Fig.\ref{fig:inversion} a. The uncertainty in pulse time seeds a starting magnetization, which implies an initial normalized emission rate of \qty{1.015\pm 0.015}, well below our experimental resolution. % and an initial magnetization of $\sqrt{S_x^2+S_y^2}/2N$=\qty{.1\pm.1}.  

The frequency of the $\qty{841}{nm}$ diode laser is locked to a high finesse ULE (ultralow expansion) cavity, with a measured laser linewidth of \qty{300}{Hz}. 

\subsection{Imaging}
We image the excited state atoms with single site resolution by first blowing out all ground state atoms in $0.02 \tau$ using the \qty{30}{MHz} \qty{401}{nm} transition, waiting for $500 \tau$ so that the excited state atoms decay to the ground state, and then imaging the ground state atoms in an expanded accordion lattice as per~\cite{su_fast_2024}. Quantifying the superradiant dynamics is very forgiving; however, subradiance is easily overestimated as the probability to misidentify a hole as a particle, $p_{g\to e}$, does not ``decay" as a function of time and therefore appears as a perfectly subradiant state. We aggressively bias the photon threshold to identify an atom, resulting in $p_{g\to e}=$\qty{.0002\pm.0002} and $p_{e\to g}=$\qty{.006} error rate. The $p_{e\to g}$ error can essentially be folded into the atom loss rate due to motional excitation (see Methods E.) as well as finite blowout time and is negligible compared to those loss rates, which are on the order of $0.14$. We do not correct for the measured $p_{g\to e}$ error, which is estimated to contribute less than a $3\%$ systematic subradiance enhancement for the most subradiant states at $t_e\Gamma=20$ in Fig. \Ref{fig:supsub} b., and is negligible everywhere else. The short $0.02\tau$ blowout duration limits the $p_{g\to e}$ error, which can be improved to essentially arbitrary levels at the expense of increasing $p_{e\to g}$.

\subsection{Decay Curve Benchmarking}
\begin{figure}[htp!]
    \includegraphics[width=0.48\textwidth]{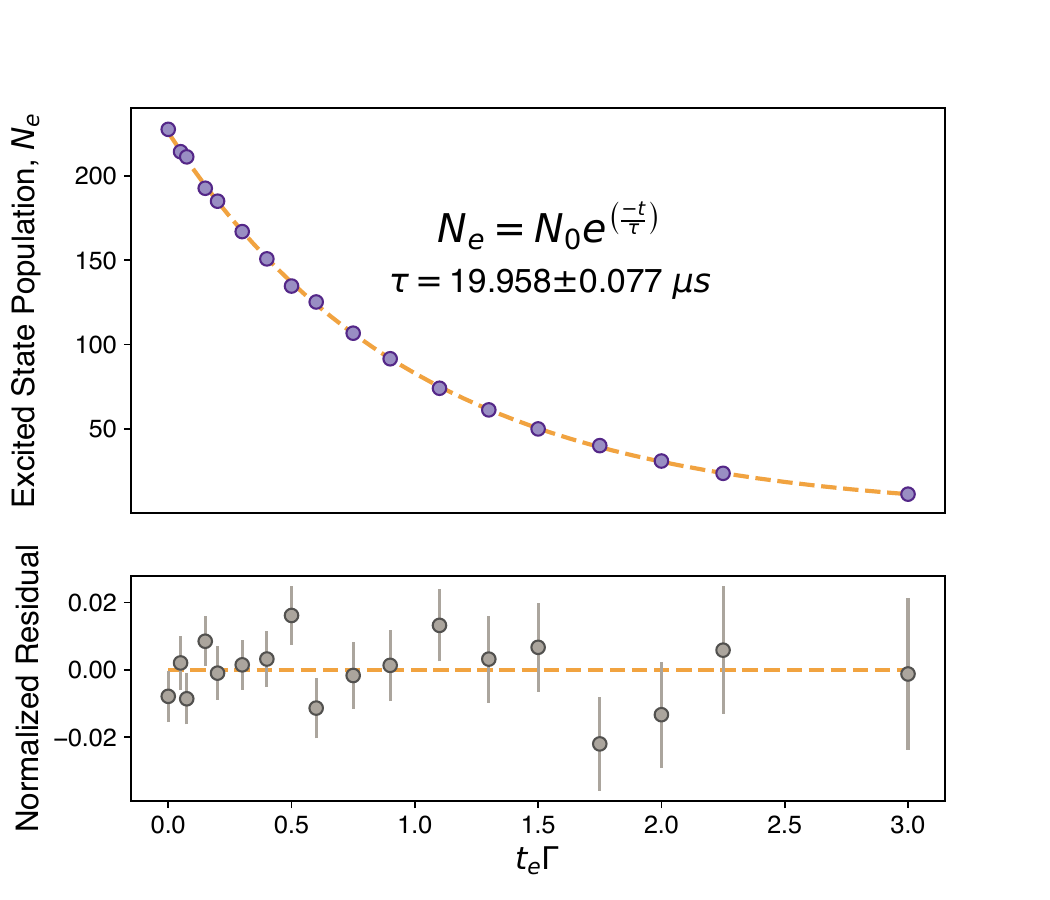}
    \caption{\textbf{Independent Decay Benchmarking}. Excited state population decay in a widely spaced lattice, $a = \qty{3}{\micro\meter}$. We fit to a simple exponential (left column) and extract the time constant $\tau$. normalized residuals are plotted below.
    }
    \label{fig:indep}
\end{figure}

To calibrate the expected decay time scale, $\tau$, and confirm that we observe cooperative effects, we measure the independent decay curve of erbium atoms in a large \qty{3}{\micro\meter} spacing lattice ($a/\lambda = 3.57$). This large spacing should effectively eliminate all cooperative effects. In Fig. \ref{fig:indep} we fit the resulting decay curve to a simple exponential $N_{e} = N_0e^{-t/\tau}$ and report an independent excited state decay lifetime of $\tau= $\qty{19.958\pm 0.077}{us} in agreement with the reported lifetime from erbium vapor decay measurements $\tau =$ \qty{20 \pm 4}{\micro\second}~\cite{ban_laser_2005}. The normalized residuals associated with this fit are plotted below and do not show any structure, as expected for simple exponential decay. For convenience, we use $\tau=$\qty{20}{\micro\second} throughout. 

\subsection{Atom Loss and Motional Excitations}
Due to the finite optical trap frequencies and different trap depths for the \qty{841}{\nm} excited state, our system experiences small amounts of atom loss during the experiment. Specifically, the finite Lamb-Dicke parameter means that the photon recoil adds about $.045$ motional excitations into the first excited band along the laser propagation direction. In addition, the different trap frequencies between the ground and excited states lead to different-sized Wannier functions. The finite overlap between these Wannier functions can result in atoms transferring into, primarily, the second excited motional state along each axis without the assistance of a recoil event, this transfer probability is listed in Table \ref{tab:traps}. For simplicity in the theory, we model this as bringing the atoms into the first motional state instead. Combining the differential polarizability and the Lamb-Dicke parameter yields a total of 0.05 motional excitations from the Rabi pulse that initialises the dynamics. Photon reemission also adds motional quanta, which we model as occurring isotropically. Due to the poorer confinement in the vertical direction, we expect 0.07 motional quanta from spontaneous emission. 

We find that, due to the accordion lattice transfer and expansion, any higher motional excitations are lost before imaging. This brings the effective $p_{e\to g}$ error rate up by $0.12$.

\subsection{Accordion Lattice}
Our tunable-spacing accordion lattice light is generated by a single-pass-doubling ytterbium-fiber-amplified laser (AzurLight Systems: ALS-BL-488-2-E-CP-SF) that produces a maximum of \qty{2}{\W} of \qty{488.1}{\nm} light. We dynamically control the spacing as well as the phase of our accordion lattice via a galvanometer and prism pair system along with an adjustable phase plate. The system generates two parallel beams of arbitrary spacing that are sent onto our in-vacuum objective to create a lattice with an arbitrary incident angle. This method of generating a lattice is heavily dependent on the characteristics of the focusing optic, in this case, our high numerical aperture (NA) objective. For an NA up to 0.85, our objective is capable of diffraction-limited performance; however, near the edges of the objective (0.85-0.92 NA), our imaging system is not diffraction-limited, resulting in decreased lattice trap depths, as can be seen in Table \ref{tab:traps} for spacings of \qty{266}{\nm}.

\subsection{Theory}

We model the setup in Fig.~\ref{fig:setup} as a collection of $N$ two-level emitters with ground state $\ket{g_i}$ and excited state $\ket{e_i}$, located at positions $\mathbf{r}_i$. All emitters have the same transition dipole moment $\mathbf{d}$ and resonance frequency $\omega_0 = c / \lambda$. The emitters couple to the common radiation field of the vacuum electromagnetic modes. Within the Born--Markov approximation, tracing out the field degrees of freedom yields the effective master equation for the emitters' density matrix $\hat{\rho}$~\cite{lehmberg_radiation_1970},
\begin{equation}
\label{eqn: MasterEquation}
    \frac{d\hat{\rho}}{dt} = - i \left[\hat{H} , \hat{\rho}\right] + \mathcal{L}(\hat{\rho}).
\end{equation}
The Hamiltonian $\hat{H}$ and Lindbladian $\mathcal{L}(\hat{\rho})$ are given by
\begin{align}
    \hat{H} &= \sum_{i=1}^N \omega_0 \hat{\sigma}_i^\dagger \hat{\sigma}_i + \sum_{i \neq j} J_{ij} \hat{\sigma}_i^\dagger \hat{\sigma}_j, \\
    \mathcal{L}(\hat{\rho}) &= \sum_{i,j=1}^N \frac{\Gamma_{ij}}{2} \left( 2 \hat{\sigma}_j \hat{\rho} \hat{\sigma}_i^\dagger - \{ \hat{\sigma}_i^\dagger \hat{\sigma}_j , \hat{\rho} \} \right),
\end{align}
where $\hat{\sigma}_i = \ket{g_i}\!\bra{e_i}$ and $\hat{\sigma}_i^\dagger = \ket{e_i}\!\bra{g_i}$ denote the lowering and raising operators of the $i$-th emitter. The Hamiltonian describes coherent photon-mediated exchange interactions, while the anti-commutator term in the Lindbladian accounts for dissipative exchange. The first term of the Lindbladian encodes the correlated loss of excitations via collective photon emission.  

The coherent $J_{ij}$ and dissipative $\Gamma_{ij}$ couplings are determined by the Green’s tensor $\mathbf{G}(\mathbf{r}, \omega)$ for a point dipole in vacuum~\cite{lehmberg_radiation_1970}, and take the form
\begin{equation}
J_{ij} - i \Gamma_{ij}/2  = -\frac{3\pi \gamma_0}{\omega_0} \, \mathbf{d}^\dagger \, \mathbf{G}(\mathbf{r}_{i} - \mathbf{r}_{j}, \omega_0) \, \mathbf{d},
\end{equation} 
where $\gamma_0$ denotes the spontaneous decay rate of a single emitter.

\subsubsection{Numerical simulation via a cumulant expansion}

A direct numerical solution of the master equation becomes intractable for systems with more than about $16$ emitters due to the exponential growth of the Hilbert space dimension---the density matrix $\hat{\rho}$ has dimension $2^N \times 2^N$~\cite{clemens_collective_2003,masson_many-body_2020}. To simulate the dynamics of ensembles of up to $450$ emitters, we instead employ an approximate method based on a cumulant expansion~\cite{rubies-bigorda_characterizing_2023,plankensteiner_quantumcumulantsjl_2022,robicheaux_beyond_2021}.  

We start by deriving the equations of motion for the expectation value of a generic system operator $\hat{O}$, 
\begin{equation}
\label{eq:Heisenberg}
    \frac{d}{dt} \langle \hat{O} \rangle = \mathrm{Tr} \left( \hat{O} \, \frac{d \hat{\rho}}{dt} \right),
\end{equation}
where $\mathrm{Tr}(\cdot)$ denotes the trace over the degrees of freedom of the emitters, and the evolution of the density matrix is given by Eq.~(\ref{eqn: MasterEquation}). We express $\hat{O}$ as a product of the single-emitter operators $\hat{\sigma}_i^{ee} = \ket{e_i}\bra{e_i}$, $\hat{\sigma}_i$, and $\hat{\sigma}_i^\dagger$. An operator $\hat{O}$ of order $\alpha$ can be expressed as a product of $\alpha$ single-emitter operators (corresponding to different emitters). For example, $\hat{\sigma}_i^\dagger \sigma_j$ is a second-order operator.

The Heisenberg equations of motion~(\ref{eq:Heisenberg}) naturally couple operators of order $\alpha$ to those of order $\alpha+1$, forming a hierarchy of equations. The exact dissipative dynamics of the system (equivalent to solving the master equation) is recovered only if the hierarchy is followed up to order $N$. To obtain an approximate solution, we truncate this hierarchy by neglecting correlations involving more than $\alpha$ emitters. This allows to express operators of order $\alpha+1$ as nonlinear combinations of lower-order operators, thereby closing the equations. For instance, a first-order cumulant expansion corresponds to neglecting all two-body correlations by setting the second-order cumulant $C_2(\hat{a},\hat{b}) = \langle \hat{a}\hat{b} \rangle - \langle \hat{a}\rangle \langle \hat{b}\rangle = 0$, which is equivalent to the mean-field approximation $\langle \hat{a}\hat{b}\rangle \approx \langle \hat{a}\rangle \langle \hat{b}\rangle$. More generally, truncating at order $\alpha$ yields a closed set of $\sim N^\alpha$ nonlinear differential equations, which defines the cumulant expansion of the equations of motion to order $\alpha$

In this work, we compare the experimental results with third-order cumulant expansions ($\alpha = 3$). For initially uncorrelated states, the equations of motion for the relevant operator averages, namely $\langle \hat{\sigma}_i^{ee} \rangle$, $\langle \hat{\sigma}_i^{ee} \hat{\sigma}_j^{ee} \rangle$, $\langle \hat{\sigma}_i^\dagger \hat{\sigma}_j \rangle$, $\langle \hat{\sigma}_i^{ee} \hat{\sigma}_j^{ee} \hat{\sigma}_q^{ee} \rangle$, and $\langle \hat{\sigma}_i^{ee} \hat{\sigma}_j^\dagger \hat{\sigma}_q \rangle$, are given in Ref.~\cite{rubies-bigorda_characterizing_2023}.

\subsubsection{Incorporating experimental factors}

To accurately model the experiment, we take into account the following factors:

\textit{(i) Distribution of atomic positions.} Although the experiment targets a specific number of atoms, the stochastic loading of the optical lattice leads to fluctuations both in the total atom number and in the site occupancy, resulting in missing atoms at random positions. These distributions are experimentally characterised by imaging the lattice immediately after loading. To account for this effect, each theoretical data point in Fig.~\ref{fig:supsub}(a,b), Fig.~\ref{fig:scaling}a and Fig.~\ref{fig:inversion} represents an average over one hundred realisations of the experimental loading procedure. The numerical data in Fig.~\ref{fig:supsub}(c,d), Fig.~\ref{fig:scaling}b and Fig.~\ref{fig:spacing} were instead obtained using square lattices with unit filling.
%The correlation data in Fig.~\ref{fig:supsub}(c,d) and the variable-spacing data in Fig.~\ref{fig:spacing} were obtained using perfect square lattices of 196 and 324 atoms, respectively, while the scaling data in Fig.~\ref{fig:scaling}b were obtained from simulations of perfect square lattices of varying sizes.

\textit{(ii) Atomic confinement.} To incorporate motional effects in the lattice potential, we average the coherent and dissipative light-induced interactions $J_{ij}$ and $\Gamma_{ij}$ over the spatial extent of the atomic motional wavefunctions. This approximation is valid in the regime where the trap oscillation period is much shorter than the characteristic timescale of the atomic evolution. We model the traps as harmonic with anisotropic frequencies specified in Table~\ref{tab:table1}, and assume the atoms occupy the motional ground state in all three directions, with a $6\%$ probability of populating the first excited state along the axis parallel to the excitation drive, treating atoms expected to be in the second motional state as in the first for convenience.

\textit{(iii) Imperfect excitation pulse.} To account for the finite accuracy of the inversion pulse—arising primarily from imperfect timing—we model each atom as being independently excited with probability $96\%$. This yields an uncorrelated initial state without coherences and populations given by $\langle \hat{\sigma}_i^{ee} \rangle = 0.96$ for all $i$, $\langle \hat{\sigma}_i^{ee} \hat{\sigma}_j^{ee} \rangle = 0.96^2$ for all $i \neq j$, and $\langle \hat{\sigma}_i^{ee} \hat{\sigma}_j^{ee} \hat{\sigma}_q^{ee} \rangle = 0.96^3$ for all $i \neq j \neq q$.

\subsection{Suppression of geometric resonances by positional disorder}

\begin{figure}[htp!]
    \includegraphics[width=0.48\textwidth]{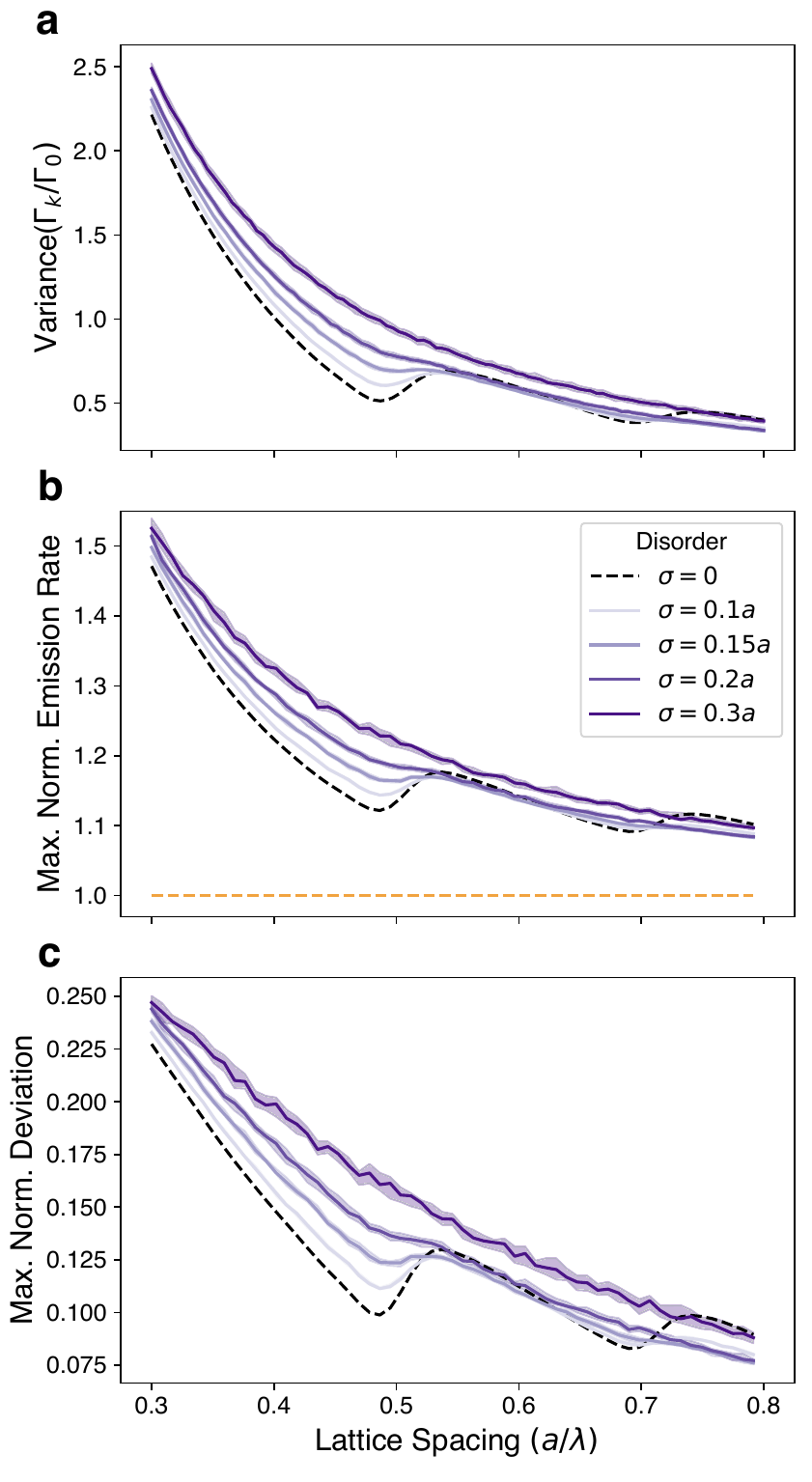}
    \caption{\textbf{Superradiant geometric resonances for disordered systems}. (a) Variance of the decay rates of the collective jump operators, (b) maximum normalized emission rate, and (c) maximum normalized deviation (that is, maximum population difference between independent and correlated decay in the superradiant regime) for $12 \times 12$ square arrays as a function of spacing $a$. We introduce positional disorder through a Gaussian distribution with zero mean and standard deviation $\sigma$. As the standard deviation is increased, the system becomes more disordered and the geometric resonances diminish and eventually vanish. For each value of $a$ and $\sigma$, we show the median and $25$-th and $75$-th percentiles of $25$ random realizations. 
    }
    \label{fig:disorder}
\end{figure}

In Section~\ref{section: variable_spacing}, we show that superradiance in atomic arrays displays a non-monotonic dependence on the lattice spacing. In particular, the maximum population difference between independent and correlated decay in the superradiant regime exhibits pronounced resonances at $a=\lambda/2$ and $a=\lambda/\sqrt{2}$. These resonances emerge when Umklapp scattering opens an additional Bragg diffraction channel, that is, when a new reciprocal lattice vector maps a Bloch momentum inside the Brillouin zone (BZ) to a momentum outside the BZ that lies within the light cone (defined as the set of Bloch momenta with magnitude smaller than the resonant photon momentum $2\pi/\lambda$).
%These resonances arise when a new reciprocal lattice vector enters the light cone (defined as the set of spin-wave momenta smaller than the resonant photon momentum $2\pi/\lambda$), thereby opening an additional Bragg diffraction channel via Umklapp scattering. 
The new diffraction order allows the corresponding atomic momentum state to couple more efficiently to free-space radiation, becoming brighter and thereby enhancing superradiance. The same resonances appear in the maximum normalized emission rate (i.e., the emission rate divided by the excited-state population), shown in Fig.~\ref{fig:disorder}b, as well as in the variance of the decay rates $\Gamma_{k}$ of the $N$ collective jump operators (where $N$ denotes the number of atoms in the array), shown in Fig.~\ref{fig:disorder}a. The latter are obtained by diagonalizing the dissipative interaction matrix with elements $\Gamma_{ij}$.

To emphasize that this effect is a genuine consequence of lattice order, we numerically study how superradiant emission is modified when positional disorder is introduced. For that, we displace the atoms from their ideal lattice sites according to a normal distribution with zero mean and standard deviation $\sigma$. As shown in Fig.~\ref{fig:disorder}, the pronounced resonances in the maximum normalized emission rate, maximum population deviation, and variance of the collective decay rates are progressively smoothed out with increasing $\sigma$, disappearing entirely once the disorder becomes sufficiently strong. This disappearance occurs because disorder destroys the lattice periodicity, eliminating the well-defined reciprocal lattice vectors needed for Bragg diffraction and the associated resonances. The same effect is seen in the decay rate of the brightest jump operator, which diminishes monotonically as disorder grows.

\subsection{Analytic Form of Independent Decay $S$ and $S_z$ Curve}
We claim that the decay curves of Dicke super radiance and independent decay are symmetric around $50\%$ of initial excitation when plotted parametrically as a function of total spin $S$ and magnetization $S_z$. The proof for Dicke super radiance is simple-as $S$ is conserved, decay is a straight line in $S_z$ space, which is necessarily symmetric. For independent decay, for an initial rotation of $\theta$ and a decay for time $t$ with rate $\gamma$, the first order observables are
\begin{align}
    \langle Z\rangle =& -1/2+\sin^2(\theta)e^{-\gamma t}\\
    \langle X\rangle =&  \cos(\theta) \sin(\theta)e^{-\gamma t/2}\\
    \langle Y\rangle = & 0
\end{align}
And the total spin are therefore:
\begin{align}
    &\sum_{i,j}\langle X_i X_j\rangle+\langle Y_i Y_j\rangle+\langle Z_i Z_j\rangle =\\
    &=3N/4+\sum_{i\neq j} (-1/2+\sin^2(\theta)e^{-\gamma t})^2+\cos(\theta) ^2\sin(\theta)^2e^{-\gamma t}\\
    &= 3N/4 + N(N-1)(1/4-e^{-2\gamma t}(e^{-\gamma t}-1)\sin^4(\theta))\\
    &= 3N/4 + N(N-1)(1/4+T(T-1)\sin^4(\theta))
\end{align}
where in the last line we have applied the transformation $t\to-\log(T)/\gamma$ with the new limits $T\in [0, 1]$. The function is symmetric around the transformation $T\to 1-T$, as desired.
\begin{figure*}
    \includegraphics[width=\textwidth]{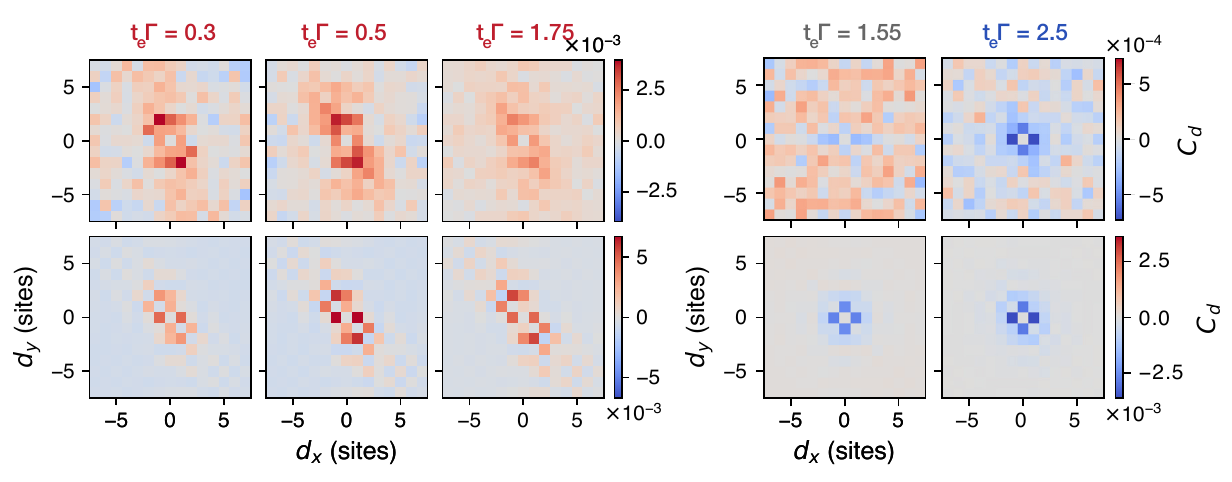}
    \caption{
    \textbf{Correlations} Correlations (from left to right) taken at $\tau\in$ [0.3, 0.5, 0.75, 1.55, 2.5], with [2200, 4000, 5000, 5000, and 5000] shots respectively. Experimental measurements (top row) and theory (bottom row) 
    }
    \label{fig:corr_supp}
\end{figure*}

\section{Data Analysis}
\subsection{Fitting Decay Traces}
Due to the collective and non-linear nature of the emission process, the excited population over time does not admit a simple fitting. At initial times $t_e=0$, the natural decay at $t_e=0$ needs to be combined with a stretched decay with an exponent greater than one to capture the superradiant dynamics at finite early times. At late times, the broad range of decay rates involved in the emission process leads to a singular stretched exponential with a low stretching exponent. These three time scales necessitate fitting to a complex sum of three stretched exponentials.

\subsection{Decay Traces and Derivatives}
Each full inversion curve for populations $[200, 350, 500, 1000]$ comprises 200 population shots taken before the decay curves to sample fidelity of the prepared Mott insulator and average number of atoms before losses due to the excitation beam, and 3960 shots for each decay curve, split between 22 time points. The instantaneous derivative is computed between two times $t_0, t_1$, with finite time difference $dt= t_1-t_0$, assuming instantaneous pure exponential decay with the same decay rate at both times, or that our measured instantaneous decay data, $D_M$, is of the form
\begin{align}
    D_M=\frac{N(t_0)-N(t_1)}{N(t_0)+N(t_1)}\frac{2}{dt}\\
    =\frac{e^{-t_0/\tau}-e^{-t_1/\tau}}{e^{-t_0/\tau}+e^{-t_1/\tau}}\frac{2}{dt}
\end{align}
Solving for the normalized decay rate at the time $t_0+dt/2)$, or halfway between the sampled points, the instantaneous normalized decay rate becomes
\begin{align*}
1/\tau=&\partial_t(e^{-t/\tau})|_{t=t_0+dt/2}/N(t_0+dt/2)\\
=&dt/(\ln(D_M dt+2)+\ln(D_M dt-2))
\end{align*}
We use this formula throughout the paper, specifically in figures \ref{fig:supsub} b. and \ref{fig:inversion} a.
In addition to the discrete derivative, we also fit the atomic population and take the numerical derivative of the fit to extract a continuous normalized emission rate. The populations in \ref{fig:supsub} and \ref{fig:inversion} are
fitted to the sum of three stretched exponentials, $N e^{-x^\alpha/\tau}$, or nine free parameters total. We observe systematics outside of experimental error bars for two or fewer stretched exponents. The functional form at any given time is expected to be a stretched exponential, where there must be a contribution of at least two different stretching exponents as the normalized derivative of a sum of normal exponents (or exponents with the same stretching factor) is always monotonically decreasing and does not exhibit a finite time peak. Furthermore, we expect several stretched exponents due to the large variance in decay rates for different excitations in the system, as well as the time-varying effective time constant for each mode due to nonlinearities between the different excitations. Note that this can be interpreted as a strongly time and momentum varying coupling to an external bath, exactly as expected of a super- and subradiant system. Error bars for all fits in the papers are computed via bootstrapping the fit with 500 samples, with shading reporting $1\sigma$ confidence intervals. Subradiance is extracted by fitting a simple exponential to the last three points of the atom traces and reporting the time constant. 

\subsection{Correlations}
Correlations were taken at times $\tau\in$ [0.3, 0.5, 0.75,1.55, 2.5], with [2200, 4000, 5000, 5000, and 5000] shots respectively and are shown in Fig \ref{fig:corr_supp} in full with theory comparison. The 5,000 shots at 1.55 $\tau$ were taken at the theoretically predicted crossover point where the correlations transition from ferromagnetic to antiferromagnetic. These results were statistically consistent with zero (as expected by theory). We focus on the center $50\%$ of sites where the Mott insulator filling is highest to minimize any correlations due to initial loading.

\subsection{Measuring Total Spin}
Fig \ref{fig:inversion} c. combines fits of the $S_z$ and $S_z^2$ measurements at finite excitation fraction in Fig \ref{fig:inversion} a. and separate measurements of the magnetization. The first points listed at $S=-0.5$ represent the expectation values for ideal Rabi pulses at the reported angles. Each other point represents experimental magnetization data, starting at \qty{4}{\micro\second} due to the finite bandwidth of the phase shifter, combined with the fits above (which have negligible errors). The magnetization is extracted as the variance of the measured excitation fraction after a $\pi/2$ pulse of phase swept from $0$ to $8\pi$. For finite inversions, we observe coherent oscillations out to $0.75\tau$; however, to avoid the possibility of laser dephasing and to include any spontaneously generated magnetization with a random phase, we measure the variance of the measured $S_z$ after the $\pi/2$ pulses rather than the contrast. As we measure the excited state population and not $S_z$ directly, any fluctuation in the initial loading fraction or atom loss artificially inflates the measured magnetization. To account for atom number variance due to initial Mott insulator loading and atom loss, we subtract off the loading variance scaled by the expected Poissonian loss $M=\sqrt{\frac{Var(N_{measured})}{2\langle N_{measured}\rangle^2}-\frac{Var(N_{loading})}{N_{measured}}}$, where M is the magnetization, $N_{measured}$ is the atom number measured during a phase sweep, and $N_{loading}$ is the loading atom number fluctuation. To further reduce the effect of loading fluctuations, we only consider the center of the Mott insulator.

\subsection{Geometric Resonances}
For each spacing in Fig \ref{fig:spacing}, we take an entire time trace, as per Fig \ref{fig:supsub}a. We then fit the first $1.75\tau$ to a sum of two stretched exponentials, $A_i e^{-(t/B_i)^{C_i}}$. Note that we do not fit the subradiant portion, and therefore do not require a third exponential. We then extract the maximum deviation from the independent decay $g$ of the fitted curve $f$, as $\text{max}_t((g(t)-f(t))/g(t))$. In fitting, we penalize $y'(0) - g'(0)$ on physical grounds. We expect a relatively large offset from theory compared to the normalized derivative in Fig \ref{fig:supsub} as the maximum deviation from independent decay is essentially the integral of the prior, and the $10\%$ theory overestimate adds up. To better compare, we fit the offset from theory across all spacings. 
%TC:endignore
\bibliography{SR_references}

\end{document}